\documentclass[aps,pre,reprint,floatfix,superscriptaddress]{revtex4-2}
\usepackage{graphicx}
\usepackage[latin1]{inputenc}
\usepackage{amsmath}
\usepackage{amsfonts}
\usepackage{amssymb}
\usepackage{xcolor}

\begin{document}
 
\title{
Classical scattering and fragmentation of clusters of ions in helical confinement
}

\date{\today}

\author{Ansgar~Siemens}
\email{asiemens@physnet.uni-hamburg.de}
\affiliation{Zentrum f\"ur Optische Quantentechnologien, Fachbereich Physik, Universit\"at Hamburg, Luruper Chaussee 149, 22761 Hamburg Germany} 
\author{Peter~Schmelcher}
\email{pschmelc@physnet.uni-hamburg.de}
\affiliation{Zentrum f\"ur Optische Quantentechnologien, Fachbereich Physik, Universit\"at Hamburg, Luruper Chaussee 149, 22761 Hamburg Germany}
\affiliation{Hamburg Center for Ultrafast Imaging, Universit\"at Hamburg, Luruper Chaussee 149, 22761 Hamburg Germany}




\begin{abstract}
\noindent We explore the scattering dynamics of classical Coulomb-interacting clusters of ions confined to a helical geometry. 
Ion clusters of equally charged particles constrained to a helix can form many-body bound states, i.e. they exhibit stable motion of Coulomb-interacting identical ions. 
We analyze the scattering and fragmentation behavior of two ion clusters, thereby understanding the rich phenomenology of their dynamics. 
The scattering dynamics is complex in the sense that it exhibits cascades of decay processes involving strongly varying cluster sizes. 
These processes are governed by the internal energy flow and the underlying oscillatory many-body potential. 
We specifically focus on the impact of the collision energy on the dynamics of individual ions during and immediately after the collision of two clusters, and on the internal dynamics of ion clusters that are excited during a cluster collision. 

\end{abstract}

\maketitle

\section{Introduction}

Wigner crystals describe a crystalline phase of Coulomb-interacting particles.
They were first predicted in 1934 by Eugene Wigner \cite{wigner1934} and can occur when the Coulomb interactions between electrons or ions dominate over their kinetic energies. 
Since their initial prediction, Wigner crystals have been detected in a large number of electronic and ionic systems \cite{goldman1990,jang2017,regan2020,grimes1979,xu2020,drewsen2015,schmoger2015,drewsen2003}, although the first direct image of a electronic Wigner crystal has been obtained only recently \cite{li2021}. 
A classical description of Wigner crystals is possible, provided the de Broglie wavelength $\Lambda$ of the ions is much smaller than the interparticle distance $a$, i.e. provided $\Lambda/a<1$. 
A simple model that captures the physics of classical Wigner crystals is the one-component plasma (OCP) \cite{baus1980}. 
It describes the statistical mechanics of classical ions in the presence of a uniformly `smeared out' background charge. 
Depending on the relative strength of Coulomb interactions and kinetic energies, the OCP exhibits gas, liquid, and Wigner crystal phases \cite{saigo2002,vieillefosse1975,hansen1973,pollock1973,ichimaru1982,schiffer2003,rogers1987}. 
While the model is a drastic oversimplification, it is used to estimate system properties that are directly impacted by strong Coulomb coupling and otherwise difficult to estimate, such as diffusion coefficients [28, 29], thermal conductivity \cite{bernu1978,donko1998,donko2004}, or shear viscosity \cite{vieillefosse1975,daligault2014}.

In finite-size systems, the properties and behavior of classical Coulomb systems depend on the trapping geometry \cite{schiffer2003}. 
There has been a strong interest in the properties of Coulomb systems that are confined to a (quasi) 1D subspace \cite{kamimura2012,hyde2013,gusein-zade2006,landa2013,dyachkov2014,piacente2004,birkl1992,bollinger1994,hasse1990}. 
The equilibrium configurations of these systems depend almost exclusively on the particle density and take the form of linear chains, zig-zag configurations, helical `chains', or multiple interwoven helical `chains' \cite{gusein-zade2005,tsytovich2005}. 
The configuration space of this system can be represented by a complex bifurcation tree that depends on the particle density within the trap \cite{landa2013}. 
Experimental realizations \cite{hyde2013,birkl1992} of these quasi-1D setups have been realized with dusty plasmas - a type of low-temperature plasma containing macroscopic-sized charged particles \cite{fortov2004,fortov2005,ignatov2005}.

Complementary to the above investigations, classical Coulomb systems have also been studied with regards to confinement along a 1D helical path \cite{schmelcher2011}. 
The combination of confining forces and the Coulomb interaction among equally charged ions gives rise to an effective interaction that allows two or more ions to form stable bound states \cite{schmelcher2011,kibis1992}. 
It should be noted that similar effective interactions in helical confinement have also been observed for excitons \cite{tokihiro1993,sarma2010} or (polarized) dipoles \cite{pedersen2014,pedersen2016,pedersen2016a,law2008}. 
Specifically the properties of helically confined ions have been investigated with regards to phonon dynamics \cite{zampetaki2015,zampetaki2015a}, the impact of external fields \cite{siemens2020,plettenberg2017,siemens2021,gloy2022}, and their extended structural crossovers \cite{zampetaki2018,zampetaki2017}. 
Specifically, it has been shown that the center-of-mass motion of ions on a helix decouples from the relative motion of the ions \cite{zampetaki2013} - providing ideal prerequisites for studying scattering dynamics. 
The scattering of few-body bound states at inhomogeneities along the helix has already demonstrated a potential for complex scattering behavior \cite{zampetaki2013}.

This work studies the scattering dynamics of many-body bound states of ions in helical confinement. 
These bound states or clusters of individual ions can become unstable during the collision, which leads to their break-up into multiple smaller ion clusters. 
These clusters frequently scatter with other clusters of bound ions, leading to a decay cascade involving multiple final state interactions. 
We identify and analyze relevant scattering regimes, thereby understanding the underlying dynamical mechanisms. 
This includes the energy-dependent collision of ion clusters and their excitation dynamics, which involves an intricate energy flow and energy redistribution process. 
Our analysis suggests the helical scattering setup as an intriguing platform for complex scattering dynamics with a plethora of intermediate state interaction processes due to the long-range Coulomb interaction in combination with the curved geometry.

This work is structured as follows. 
Section \ref{Sec:General_Setup} introduces the mathematical model and explains the basic properties of ions in helical confinement. 
It also provides an overview of the most common features occurring during a scattering event and introduces the relevant nomenclature. 
The phenomenology of the collision dynamics of two ion clusters is described in detail in Sec. \ref{Sec:initialCollision}. 
The excitation and subsequent break-up of clusters is investigated in Sec. \ref{Sec:breakUp}. 
Finally, Section \ref{Sec:Summary} provides a brief summary and concluding remarks.

\section{Scattering of ions in helical confinement}
\label{Sec:General_Setup}

\begin{figure}
	\includegraphics[width=\columnwidth]{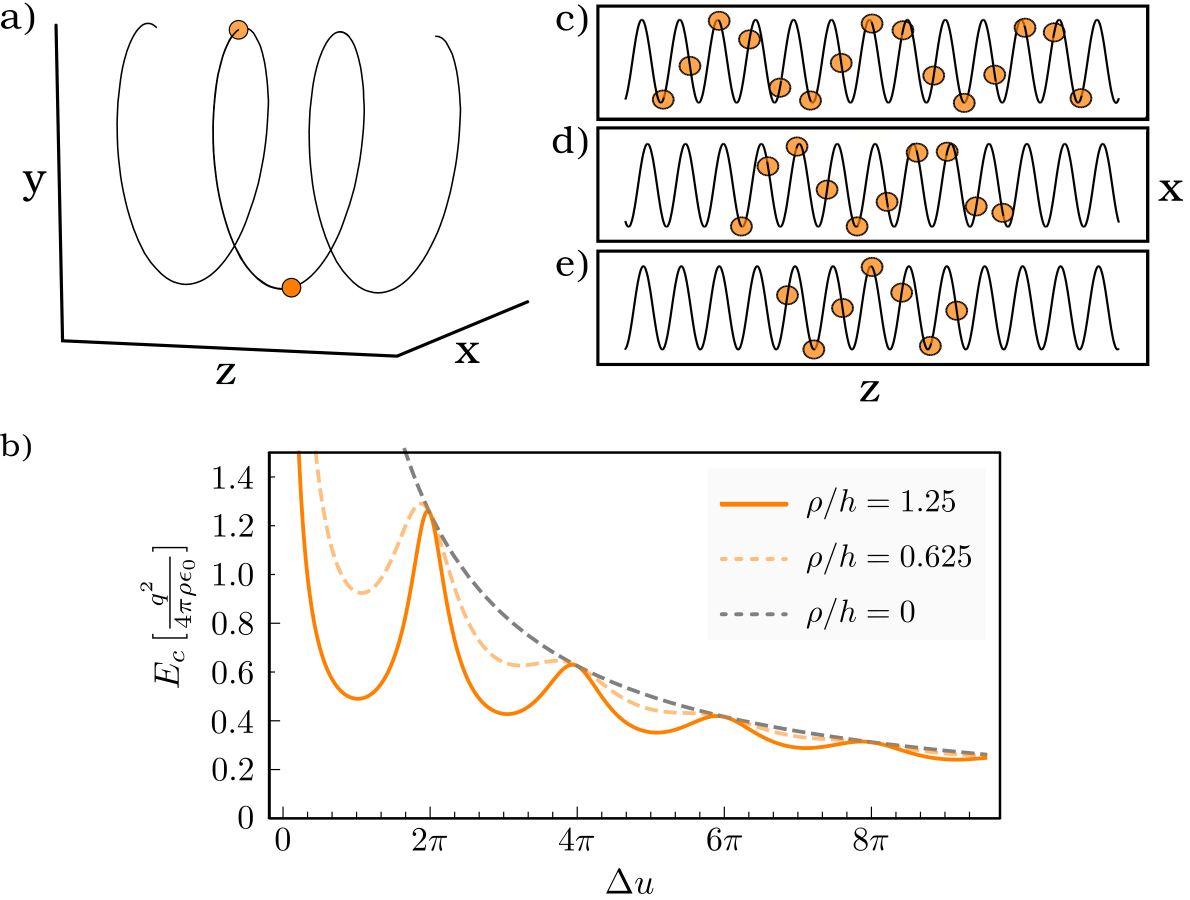}
	\caption{\label{Fig:equilibria} 
		(a) Equilibrium configuration of two ions (orange) on a helix with $\rho/h=1.25$. 
		(b) Effective interaction potential for two ions for different values of $\rho/h$. 
		(c-e) Sketches for equilibrium configurations used for the scattering simulations consisting of $N=15$, $10$, and $7$ ions in terms of the $(x,z)$-projection of the helical setup. 
	}
\end{figure}

\paragraph*{Setup and binding.}
We consider ions with an equal charge $q$ confined to move along a helix. 
The `allowed' position of the ions along the helical path can be parameterized using the following parametric equation
\begin{equation}
	\textbf{r}(u)=
	\left(
	\begin{array}{c}
		x(u) \\
		y(u) \\
		z(u)
	\end{array}
	\right) = 
		\left(
	\begin{array}{c}
		\rho\cos(u) \\
		\rho\sin(u) \\
		hu/2\pi
	\end{array}
	\right), 
\end{equation}
where $\rho$ and $h$ are the helix radius and pitch, respectively. 
A visualization of such a helix is shown in Fig. \ref{Fig:equilibria}(a). 
We use the following nomenclature: the position of the $i$-th ion in the system is given by a parametric coordinate $u_i$ corresponding to the position $\textbf{r}_i = \textbf{r}(u_i)$ in Euclidean, i.e. 3D, space. 
The ions interact via Coulomb interaction. 
Consequently, the Lagrangian of a system of $N$ ions with equal charge $q$ and mass $m$ is given by 
\begin{equation}\label{eq:helixLagrange}
	\begin{aligned}
		&\mathcal{L}(u_1,u_2,...u_N) \\ 
		&= \sum_{i=1}^{N} \dfrac{m}{2} \dot{\textbf{r}}^2(u_i) -  \sum_{\substack{i,k=1 \\ k>i}}^{N} 	\frac{1}{4\pi\epsilon_0}\frac{q^2}{|\textbf{r}(u_i) - \textbf{r}(u_k)|}  \\
		=& \sum_{i=1}^{N} \xi \dot{u}_i^2  - \sum_{\substack{i,k=1 \\ k>i}}^{N} k \left(2\rho^2\left(1-\cos(\Delta_{ik})\right) + \left(\dfrac{h\Delta_{ik}}{2\pi}\right)^2 \right)^{-\frac{1}{2}},
	\end{aligned}
\end{equation}
where $u_i$ is the position of the $i$-th particle in parametric coordinates, $\Delta_{ik} = (u_i - u_k)$ is the (parametric) distance of the $i$-th and $k$-th particle, $k=q^2/4\pi\epsilon_0$ is the Coulomb constant, and $\xi = m|\partial \textbf{r}(u)/\partial u|^2 /2= m \left(\rho^2 + \left(h/2\pi\right)^2\right)/2$ is the particles effective mass. 
Especially notable is the potential term: 
Instead of being purely repulsive, the pairwise potential energy between the ions consists of a denominator with two (very different!) components - one depending on the helix radius $\rho$ and the other depending on the helix pitch $h$. 
The $h$-dependent term is linear in $\Delta_{ik}$, whereas the $\rho$-dependent term oscillates with the parametric distance $\Delta_{ik}$. 
For a vanishing helix radius $\rho\rightarrow0$, i.e., the limit of a straight line, the potential simplifies to the standard Coulomb potential. 
When the helix pitch vanishes $h\rightarrow0$, i.e., the limit of a circle, the Coulomb potential becomes periodic, and has a minimum when the two ions are located on opposite sides of the circle. 
For finite $\rho$ and $h$, the effective interaction potential between two ions is determined by the competition between these two terms. 
An example for the potential energy as the function of the distance of two ions is depicted in Fig. \ref{Fig:equilibria}(b). 
While for large distances $\Delta_{ik}$ the repulsive term dominates, the shorter range behavior is dominated by the oscillating term. 
The number and depth of the emerging potential wells can be tuned by varying the helix parameters $\rho$ and $h$. 
Each potential well corresponds to a configuration, where the two ions are trapping each other on opposite sides of the helix, being unable to move in either direction along the helical path without decreasing their Euclidean distance with respect to each other and therefore increasing the corresponding potential energy.

These oscillating effective interactions allow for the formation of bound states consisting of two or more ions. 
The potential landscape for a system with many ions can become quite complex, allowing for a plethora of possible equilibrium configurations. 
These bound states of ions are from now on referred to as clusters. 
Already for three ions, these equilibria can be quite hard to predict \cite{schmelcher2011}.

A final comment regarding the numerical calculations: 
The ion cluster dynamics can show sensitivity to initial conditions. 
We verified the convergence of our (finite time) numerical calculations with respect to changes of the time step width. 
Furthermore, we verified the conservation of the total energy throughout simulations.

\paragraph*{Scattering state.}
It has been shown that the helical center of mass motion of ions of a cluster decouples from their relative motion \cite{zampetaki2013}. 
This allows us to systematically analyze the scattering dynamics of two clusters. 
The setup of this scattering simulation is as follows: 
Two clusters are placed far apart on the helix, such that the Coulomb interactions between them is negligible. 
We will denote these two clusters as cluster A and cluster B, respectively consisting of $N_a$ and $N_b$ particles. 
At the beginning of a simulation, the ions in cluster A and B are initialized with the following momentum
\begin{equation}\label{eq:initialMomenta}
	p_A = -\sqrt{\frac{m E_c N_B}{N_A\left[N_A+N_B\right]}} \ \ \text{and} \ \ p_B = \sqrt{\frac{m E_c N_A}{N_B\left[N_A+N_B\right]}}, 
\end{equation} 
where $E_c$ is the total kinetic energy at which the clusters collide, and $m$ is the ion mass. 
The momentum is chosen such that the total momentum during a collision is conserved, i.e.  $N_A p_A + N_B p_B = 0$. 
Due to the large number of possible equilibrium configurations, we will - for simplicity - only consider a special subset of these equilibrium configurations where the ions are (insofar as possible) spaced equidistant in the parametric coordinates. 
Some examples for such configurations are visualized in Fig. \ref{Fig:equilibria}(c)-(e).

\begin{figure}
	\includegraphics[width=\columnwidth]{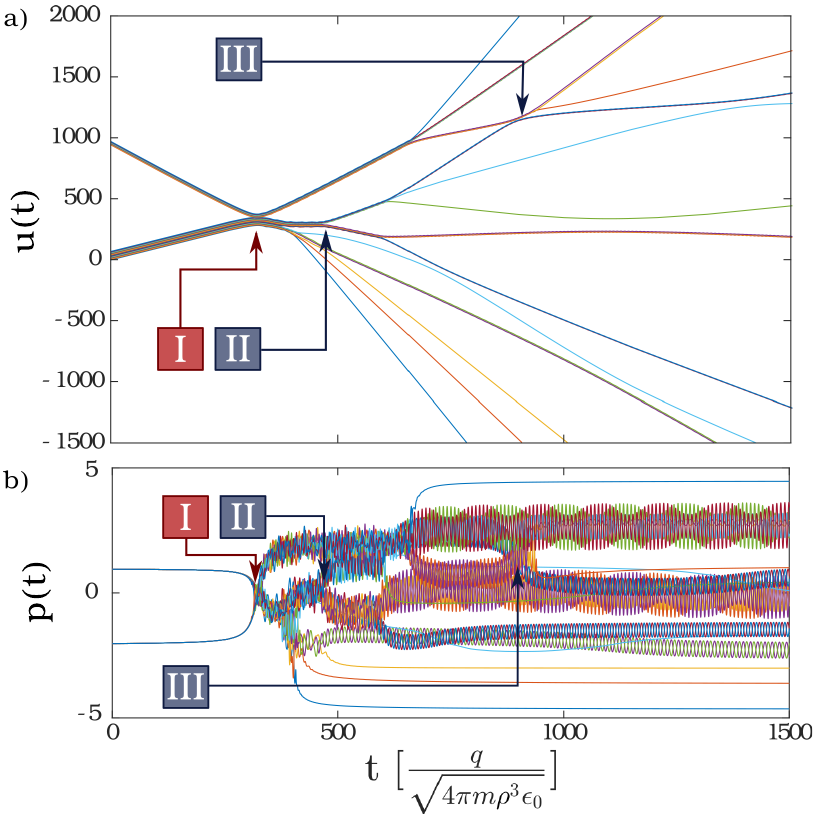}
	\caption{\label{Fig:example1} 
		Example scattering event of two clusters with $N_a=7$, $N_b=15$, $\rho/h=1.25$, and $E_c\approx21.1257 \, q^2/4\pi\rho\epsilon_0$. 
		(a) and (b) respectively show the time evolution of the (parametric) position $u(t)$ and the corresponding momentum $p(t)$ for all particles. For reference the initial collision, as well as two subsequent cluster collisions are marked in both figures. 
	}
\end{figure}

An example for such a scattering event is shown in Fig. \ref{Fig:example1}(a) for two clusters with $N_A = 7$ and $N_B = 11$ particles. 
The purpose of this example is to showcase some of the features occurring during a common collision event, before we will discuss them in more detail in later sections. 
Figure \ref{Fig:example1}(a) shows the particle positions (in parametric coordinates) as a function of time. 
Let us introduce our use of nomenclature. 
We will refer to the entire time evolution (from the initial $t=0$ until a `stable' asymptotic final state is reached) as a scattering event. 
The scattering event shown in Fig. \ref{Fig:example1} is made up of several sub-collisions or cluster break-ups, some of which are indicated by roman numerals. 
We will refer to sub-collisions, such as those marked I and III in Fig. \ref{Fig:example1}(a), as cluster collisions. 
In general, the very first cluster collision is a special case and will from here on be referred to as the initial collision. 
In the example in Fig. \ref{Fig:example1}, the initial collision is indicated by the numeral I. 
To analyze the dynamics of the individual ions during a scattering event, the corresponding momentum time evolution of all particles is shown in Fig. \ref{Fig:example1}(b). 
The numerals were added to the figure to simplify the mapping from position to momentum space. 
Before the initial collision, particles in the clusters respectively move with momentum $p_A$ and $p_B$ (as defined in Eq. \ref{eq:initialMomenta}). 
During the initial collision, the ions will slow down. 
The increasing Coulomb interaction between the two clusters will excite the internal dynamics of the ions within a cluster. 
In the figure, these excitations are recognizable as rapid oscillations of the ions momenta as the two clusters drift apart, following e.g. the initial collision I. 
As a result of this kinetic energy in the relative motion of the ions, the clusters become unstable and will break up some time after the initial collision. 
One example for a cluster break up is indicated as II in Fig. \ref{Fig:example1}. 
The ions and clusters resulting from such a breakup may further collide with other ions or break up clusters, or they might break up themselves further into even smaller clusters. 
The stability and the break-up dynamics of clusters are discussed in more detail in Sec. \ref{Sec:breakUp}. 
These subsequent cluster collisions are reminiscent of the initial collision - with the major difference being that the clusters during these secondary (and higher) collisions are excited ones and may not be stable configurations. 
More details on the dynamics during (and immediately after) a cluster collision are provided in Sec. \ref{Sec:initialCollision}.

\section{Cluster collisions}
\label{Sec:initialCollision}

Scattering events can be roughly classified into three scattering regimes - the regimes of low, medium, and large scattering energies. 
If the collision energy is too low, the two clusters never come significantly close to each other. 
Since at these large distances the repulsive part of the effective Coulomb interaction (Eq. \ref{eq:helixLagrange}) dominates over the oscillating part, the two clusters will simply slow down and then recede. 
During such `collisions' in the low-energy regime, the amount of energy that is converted into excitations of clusters internal degrees of freedom is typically very small. 
We will therefore refer to this low-energy scattering regime as the elastic regime.  
On the other hand, if the collision energy is very large, the ions will be too fast to significantly interact with one another. 
In this limit the interaction effectively reduces to a ballistic contact-type interaction. 
However, in the regime of intermediate collision energies, the oscillating effective interaction potential becomes relevant for the scattering dynamics. 
The example shown in Fig. \ref{Fig:example1} is from this intermediate regime. 
Scattering events from this intermediate regime generally exhibit multiple cluster collisions and cluster break ups. 
We will therefore refer to this intermediate regime as the inelastic scattering regime.

\begin{figure}
	\includegraphics[width=\columnwidth]{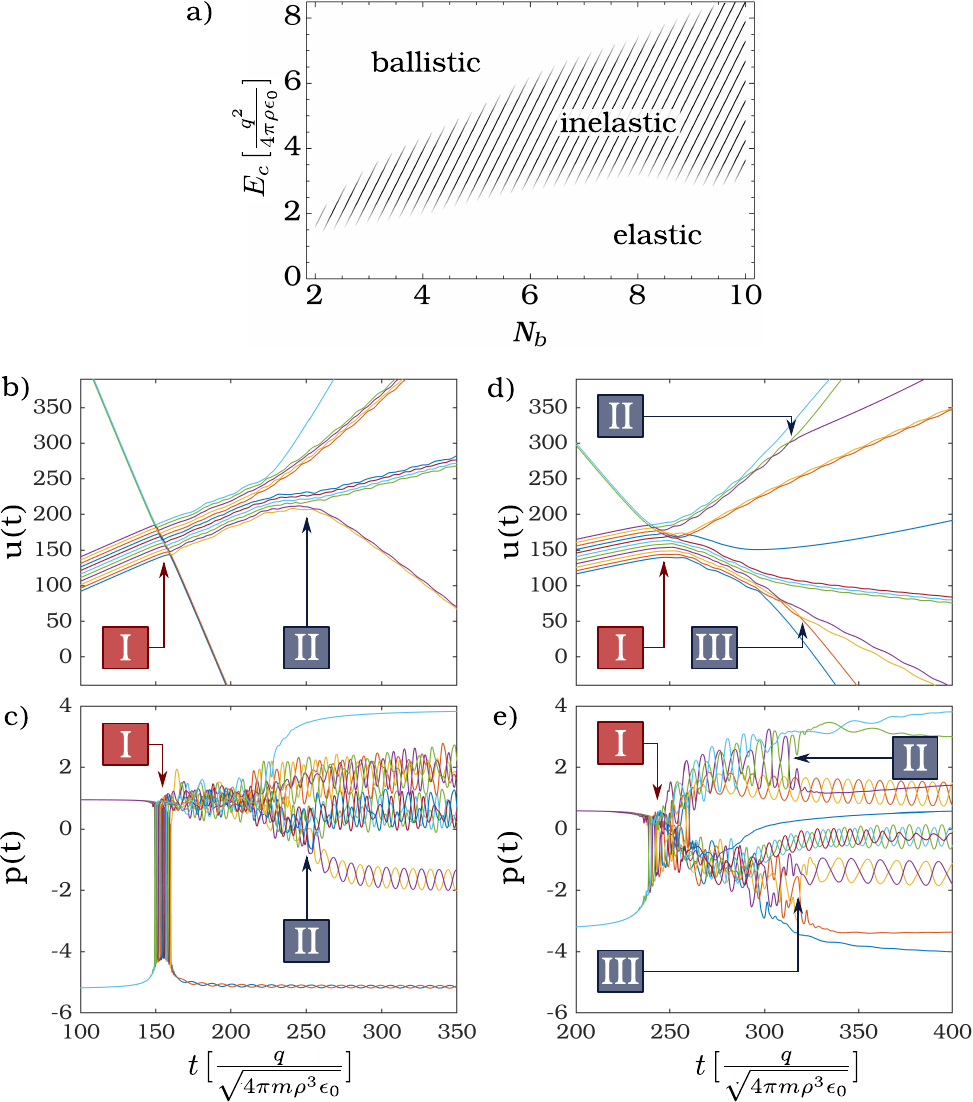}
	\caption{\label{Fig:example2} 
		(a) Scattering regimes for collisions of a single particle ($N_A=1$) with a cluster of $N_B$ ions on a helix with $\rho/h=1.25$. 
		(b-e) Scattering events with $N_A=2$ and $N_B=11$ in parametric coordinates and momentum space. 
		The collision energies are (b-c) $E_c\approx29.3773 \, q^2/4\pi\rho\epsilon_0$ and (d-e) $E_c\approx12.9559 \, q^2/4\pi\rho\epsilon_0$. 
		The numerals I-III were added to (b-e) to simplify the mapping between position and momentum space. 
	}
\end{figure}

The collision energies that mark the boundaries between the scattering regimes can be determined for scattering events with given particle numbers $N_A$ and $N_B$. 
As a sketch of the overall behavior, Fig. \ref{Fig:example2}(a) shows these regime boundaries for the case of a single ion colliding with clusters containing up to $N_b=10$ ions. 
Note, that the shown regime boundaries are fuzzy and in the present context the inelastic scattering regime is characterized by the breakup of clusters. 
Therefore, the regime boundaries in Fig. \ref{Fig:example2}(a) are determined as the lowest and largest collision energies $E_c$ for which we observe such breakups.

We now analyze the dynamics during the initial cluster collision, with a focus on the transition from the elastic to the ballistic scattering regime. 
Examples for scattering events from the inelastic regime are shown in Fig. \ref{Fig:example1} and Figs. \ref{Fig:example2}(b)-(e). 
The inelastic scattering  event shown in Fig. \ref{Fig:example1} is, as compared to Fig. \ref{Fig:example2}, somewhat close to the elastic regime. 
There, the two clusters will mainly invert their momentum during the initial collision (see the cluster collision marked I in Fig. \ref{Fig:example1}). 
Only a comparatively small part of the collision energy is transferred to excite the internal dynamics of ions within a cluster. 
Breakups of clusters occur some time after the initial collision, see for example the point marked II in Fig. \ref{Fig:example1}. 
An example for an inelastic scattering event closer to the ballistic regime is shown in Figs. \ref{Fig:example2}(b)-(c). 
Also in this example, only a comparatively small part of the collision energy is transferred to excite the internal dynamics of ions within a cluster. 
In the initial collision, the ions exchange momentum, such that the number of particles with momentum $p\approx p_A$ and $p\approx p_B$ before and after the collision is conserved. 
Breakups of clusters also occur here some time after the initial collision event [see e.g. marker II in Fig. \ref{Fig:example2}(b)]. 
An example for a scattering event deep in the inelastic regime is shown in Figs. \ref{Fig:example2}(d-e). 
In this example, there is a significant energy transfer during the initial collision of the clusters. 
As a result of this energy transfer, the clusters break up already during the initial collision [see I in Fig. \ref{Fig:example2}(d)]. 
Further breakups or cluster collisions can be observed some time after the initial collision [see e.g. II and III in Fig. \ref{Fig:example2}(d-e)]. 
For most scattering events in the inelastic regime, clusters will break up during the initial collision, akin to the example shown in Figs. \ref{Fig:example2}(d-e). 
Scattering events from the inelastic regime for which the clusters do not break up during the initial collision are typically only observed for very large or very small scattering energies, i.e. close to the ballistic or close to the elastic regime.

A comment on the break-up of clusters is in order. 
It was already discussed that clusters can be excited with some internal energy and become unstable, leading to their breakup. 
In that regard, there is one other observation that can be made from the above-discussed example events. 
A close examination of the momentum time evolution indicates that stable clusters predominantly exhibit periodic motion [see e.g. the momentum time evolution following the cluster breakup marked III in Fig. \ref{Fig:example1}(b)]. 
In contrast, the dynamics of unstable clusters is characterized by a non-periodic high-frequency time evolution [see the momentum time evolution following e.g. I in Fig. \ref{Fig:example2}(c) or II in Fig. \ref{Fig:example1}(b)]. 
The dynamics of excited and unstable clusters is discussed in Sec. \ref{Sec:breakUp} below.

\section{On the excitation and decay of clusters}\label{Sec:breakUp}

\begin{figure}
	\includegraphics[width=\columnwidth]{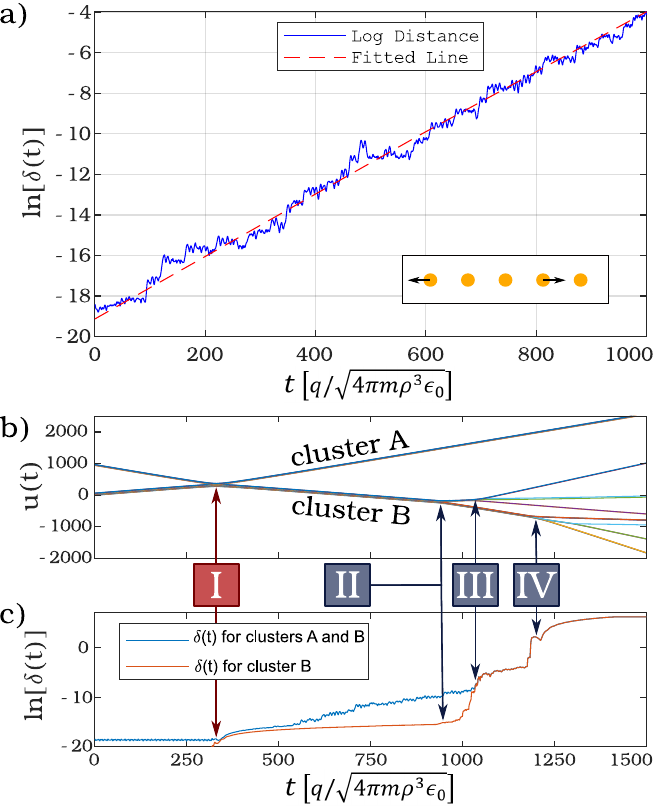}
	\caption{\label{Fig:decay} 
		(a) Time evolution of the phase space distance $\delta(t)$ for two trajectories with $\delta(0)=10^{-8}$ and $E_{ex}\approx0.2411$. See text for details. 
		The inset is a sketch that visualizes the initial excitation of ions at $t=0$. 
		(b) Parametric positions $u(t)$ of the ions for an example collision with $N_A=7$, $N_B=15$, $\rho/h=1.25$, and $E_c\approx20.4747 \, q^2/4\pi\rho\epsilon_0$. 
		(c) The blue curve shows the phase-space distance $\delta(t)$ between the trajectory shown in (b) and a reference trajectory with an initial distance of $\delta(0) = 10^{-8}$. The contribution of the $N_B=15$ particles from cluster B to $\delta(t)$ is shown in orange. 
	}
\end{figure}

\paragraph*{Internal dynamics.}
We now investigate the internal dynamics of ions within a cluster by analyzing a specific example case. 
We consider a single cluster consisting of $N=5$ particles. 
The cluster is excited at time $t=0$ with an energy $E_{ex}$. 
This excitation corresponds to `kicking' two particles (namely the first and the second to last particle) of the cluster with equal strength into opposite directions. 
A sketch of this excitation is provided in the inset of Fig. \ref{Fig:decay}(a). 
It is also helpful for the analysis to introduce the (phase-space) trajectory $\textbf{S}(t)$of the cluster, which is defined as 
\begin{equation}
	\textbf{S}(t)=(u_1(t),...u_N(t),p_1(t), ...p_N(t)). 
\end{equation}

The dynamics of this excited cluster can show sensitivity to initial conditions. 
This can be analyzed by considering the time evolution of the phase-space distance $\delta(t)$ between two trajectories $\textbf{S}(t)$ and $\textbf{S}_{ref}(t)$, where $\delta(t)$ is defined as
\begin{equation}
	\delta(t) = \sqrt{\left[\textbf{S}(t)-\textbf{S}_{ref}(t)\right]^2}.  
\end{equation}
In the following, we consider trajectories $\textbf{S}(t)$ and  $\textbf{S}_{ref}(t)$ that are initially separated by $\delta(0)=10^{-8}$. 
In this case, the sensitivity to initial conditions manifests as an exponential growth of $\delta(t)$. 
An example for this is shown in Fig. \ref{Fig:decay}(a) for an excitation with $E_{ex}=0.491 q^2/4\pi\epsilon_0\rho$ . 
The red dashed line in Fig. \ref{Fig:decay}(a) is a fit function that highlights the net exponential growth of $\delta(t)$. 
Note that in general this exponential growth of $\delta(t)$ only persists for a finite time: 
Either the cluster decays, or $\delta(t)$ reaches a maximum that depends on the size of the (energetically) accessible phase space region. 
There are two limiting cases: 
In the limiting case of low excitation energies $E_{ex}$, corresponding to weak kicks, the dynamics become linear - resulting in $\delta(t)\approx10^{-8}=const.$. 
In the opposite limit of large excitation energies $E_{ex}$, corresponding to strong kicks, the cluster breaks-up almost immediately - long before $\delta(t)$ increases to a significant value.

\paragraph*{Scattering events.}
Scattering events can show strong linear growth of $\delta(t)$, i.e. a logarithmic growth of $\ln[\delta(t)]$. 
An example that demonstrates this is shown in Fig. \ref{Fig:decay}(b)-(c). 
Figure \ref{Fig:decay}(b) shows particle trajectories $u(t)$ for a scattering event with two clusters consisting of $N_A=7$ and $N_B=15$ particles. 
These two clusters are from here on referred to as cluster A (close to $u=1000$ at $t=0$) and cluster B (close to $u=0$ at $t=0$) respectively. 
Just like before, we examine the time evolution of the phase space distance $\delta(t)$ with respect to a reference trajectory that is initially separated by $\delta(0)=10^{-8}$. 
This is shown by the blue curve in Fig. \ref{Fig:decay}(c). 
The orange curve in Fig. \ref{Fig:decay}(c) shows the contribution of cluster B to $\delta(t)$. 
In accordance with the figure, we will from now on describe $\ln[\delta(t)]$ rather than $\delta(t)$. 
Between $t=0$ and the initial collision [I in Fig. \ref{Fig:decay}(c)], the clusters do not significantly interact and all kinetic energy is in the center-of-mass motion. 
During this time, $\ln[\delta(t)]$ remains approximately constant. 
There is a sharp logarithmic increase of $\ln[\delta(t)]$ immediately following the initial collision [marked I in Fig. \ref{Fig:decay}(b-c)]. 
Only some time after the initial collision (around $t\sim500$) does the growth of $\ln[\delta(t)]$ become linear. 
Logarithmic growth of $\ln[\delta(t)]$ can also be observed following the cluster breakups marked II and IV in Figs. \ref{Fig:decay}(b-c).

This logarithmic increase of $\ln[\delta(t)]$ can be explained with the center of mass (CoM) motion of the clusters: 
The CoM of each cluster can be assigned a corresponding CoM momentum. 
When the CoM momentum of a cluster is different from the corresponding CoM momentum of the reference trajectory, $\delta(t)$ will grow linearly - corresponding to logarithmic growth of $\ln[\delta(t)]$. 
In this case, the difference of CoM momenta of clusters masks the exponential divergence due to the internal dynamics. 
Exponential growth of $\delta(t)$, i.e. linear growth of $\ln[\delta(t)]$, requires the differences in CoM momentum to be sufficiently small, such that the dominant contribution to $\delta(t)$ stems from differences in the relative motion of ions within a cluster.

The behavior of $\delta(t)$ for the example shown in Fig. \ref{Fig:decay}(c) is of general character: 
Cluster collisions and cluster breakups usually result in logarithmic growth of $\ln[\delta(t)]$. 
Linear growth of $\ln[\delta(t)]$ can be observed - but it typically occurs only for finite times. 
For small system sizes ($<30$ ions), small changes in the initial conditions typically do not lead to significant changes in the ion trajectories. 
For example, on the overall scale of the event, the trajectory shown in Fig. \ref{Fig:decay}(b) shows no discernible differences from the reference trajectory that was used to determine $\delta(t)$ - with $\delta(t)/N_A+N_B<2.5$ for $0<t<1500$. 
However, discernible differences between initially nearby (phase-space) trajectories can be expected to occur for larger system sizes (with $\gg30$ ions).

\section{Brief summary and conclusions}\label{Sec:Summary}

We have analyzed the scattering and intricate multiple break up dynamics of bound ion clusters in a helical geometry. 
During collisions of clusters, energy can be transferred to the internal motion of ions of subclusters. 
Strong excitations of these internal degrees of freedom can result in clusters becoming unstable. 
The breakup of unstable clusters leads to secondary, tertiary, and even higher order cluster collisions, such that the complete scattering process can take a very long time to reach a stable asymptotic final state.

We classified scattering events into three scattering regimes: an elastic, inelastic, and ballistic scattering regime. 
For scattering events in the inelastic regime, the dynamics of cluster collisions and their dependence on the collision energy was analyzed and described. 
For the dynamics of ions within a cluster we demonstrated a sensitivity of these dynamics to initial conditions that is characterized by an exponential divergence of initially nearby phase-space trajectories. 
We further showed, that cluster collisions and breakups typically result in a linear divergence of nearby phase-space trajectories.

The presented examples have all considered relatively small particle numbers, with $N_A+N_B<30$. 
One reason for that is, that scattering events quickly increase in complexity when the particle number increases, as a result of which inspecting and analyzing the collision dynamics becomes difficult. 
Nevertheless, a final comment on the extent to which the discussed behavior can be generalized for larger systems is in order. 
Overall, we did not observe any qualitative differences in the scattering dynamics with increasing system size. 
The transition between elastic and ballistic scattering regimes with increasing collision energy, as well as the dynamics of excited clusters behave the same. 
The stability of clusters generally decreases with increasing particle number - making large clusters more volatile. 
However, this decreased stability can be counteracted by changing the helical geometry, i.e. by increasing $\rho/h$.

Experimental realizations of our setup could be envisioned in the context of cold atoms. 
Helical trapping potentials have already been realized for neutral atoms \cite{reitz2012}. 
Furthermore, the investigated regime of $<30$ ions is certainly accessible with current trapping techniques for cold ions.  
For example, linear Paul traps can be used for the controlled capture of small particle numbers, ranging from single ions up to larger ensembles of hundreds or even thousands of ions \cite{willitsch2012}.
Finally, on a more speculative side, the here-observed cluster dynamics are (purely phenomenologically) reminiscent of heavy ion collisions.

\begin{acknowledgments}
The authors acknowledge inspiring discussions with F.K. Diakonos.
\end{acknowledgments}


\bibliography{txtest}

\begin{thebibliography}{55}%
\makeatletter
\providecommand \@ifxundefined [1]{%
 \@ifx{#1\undefined}
}%
\providecommand \@ifnum [1]{%
 \ifnum #1\expandafter \@firstoftwo
 \else \expandafter \@secondoftwo
 \fi
}%
\providecommand \@ifx [1]{%
 \ifx #1\expandafter \@firstoftwo
 \else \expandafter \@secondoftwo
 \fi
}%
\providecommand \natexlab [1]{#1}%
\providecommand \enquote  [1]{``#1''}%
\providecommand \bibnamefont  [1]{#1}%
\providecommand \bibfnamefont [1]{#1}%
\providecommand \citenamefont [1]{#1}%
\providecommand \href@noop [0]{\@secondoftwo}%
\providecommand \href [0]{\begingroup \@sanitize@url \@href}%
\providecommand \@href[1]{\@@startlink{#1}\@@href}%
\providecommand \@@href[1]{\endgroup#1\@@endlink}%
\providecommand \@sanitize@url [0]{\catcode `\\12\catcode `\$12\catcode
  `\&12\catcode `\#12\catcode `\^12\catcode `\_12\catcode `\%12\relax}%
\providecommand \@@startlink[1]{}%
\providecommand \@@endlink[0]{}%
\providecommand \url  [0]{\begingroup\@sanitize@url \@url }%
\providecommand \@url [1]{\endgroup\@href {#1}{\urlprefix }}%
\providecommand \urlprefix  [0]{URL }%
\providecommand \Eprint [0]{\href }%
\providecommand \doibase [0]{https://doi.org/}%
\providecommand \selectlanguage [0]{\@gobble}%
\providecommand \bibinfo  [0]{\@secondoftwo}%
\providecommand \bibfield  [0]{\@secondoftwo}%
\providecommand \translation [1]{[#1]}%
\providecommand \BibitemOpen [0]{}%
\providecommand \bibitemStop [0]{}%
\providecommand \bibitemNoStop [0]{.\EOS\space}%
\providecommand \EOS [0]{\spacefactor3000\relax}%
\providecommand \BibitemShut  [1]{\csname bibitem#1\endcsname}%
\let\auto@bib@innerbib\@empty
\bibitem [{\citenamefont {Wigner}(1934)}]{wigner1934}%
  \BibitemOpen
  \bibfield  {author} {\bibinfo {author} {\bibfnamefont {E.}~\bibnamefont
  {Wigner}},\ }\bibfield  {title} {\bibinfo {title} {On the {{Interaction}} of
  {{Electrons}} in {{Metals}}},\ }\href
  {https://doi.org/10.1103/PhysRev.46.1002} {\bibfield  {journal} {\bibinfo
  {journal} {Phys. Rev.}\ }\textbf {\bibinfo {volume} {46}},\ \bibinfo {pages}
  {1002} (\bibinfo {year} {1934})}\BibitemShut {NoStop}%
\bibitem [{\citenamefont {Goldman}\ \emph {et~al.}(1990)\citenamefont
  {Goldman}, \citenamefont {Santos}, \citenamefont {Shayegan},\ and\
  \citenamefont {Cunningham}}]{goldman1990}%
  \BibitemOpen
  \bibfield  {author} {\bibinfo {author} {\bibfnamefont {V.~J.}\ \bibnamefont
  {Goldman}}, \bibinfo {author} {\bibfnamefont {M.}~\bibnamefont {Santos}},
  \bibinfo {author} {\bibfnamefont {M.}~\bibnamefont {Shayegan}},\ and\
  \bibinfo {author} {\bibfnamefont {J.~E.}\ \bibnamefont {Cunningham}},\
  }\bibfield  {title} {\bibinfo {title} {Evidence for two-dimentional quantum
  {{Wigner}} crystal},\ }\href {https://doi.org/10.1103/PhysRevLett.65.2189}
  {\bibfield  {journal} {\bibinfo  {journal} {Phys. Rev. Lett.}\ }\textbf
  {\bibinfo {volume} {65}},\ \bibinfo {pages} {2189} (\bibinfo {year}
  {1990})}\BibitemShut {NoStop}%
\bibitem [{\citenamefont {Jang}\ \emph {et~al.}(2017)\citenamefont {Jang},
  \citenamefont {Hunt}, \citenamefont {Pfeiffer}, \citenamefont {West},\ and\
  \citenamefont {Ashoori}}]{jang2017}%
  \BibitemOpen
  \bibfield  {author} {\bibinfo {author} {\bibfnamefont {J.}~\bibnamefont
  {Jang}}, \bibinfo {author} {\bibfnamefont {B.~M.}\ \bibnamefont {Hunt}},
  \bibinfo {author} {\bibfnamefont {L.~N.}\ \bibnamefont {Pfeiffer}}, \bibinfo
  {author} {\bibfnamefont {K.~W.}\ \bibnamefont {West}},\ and\ \bibinfo
  {author} {\bibfnamefont {R.~C.}\ \bibnamefont {Ashoori}},\ }\bibfield
  {title} {\bibinfo {title} {Sharp tunnelling resonance from the vibrations of
  an electronic {{Wigner}} crystal},\ }\href
  {https://doi.org/10.1038/nphys3979} {\bibfield  {journal} {\bibinfo
  {journal} {Nature Phys.}\ }\textbf {\bibinfo {volume} {13}},\ \bibinfo
  {pages} {340} (\bibinfo {year} {2017})}\BibitemShut {NoStop}%
\bibitem [{\citenamefont {Regan}\ \emph {et~al.}(2020)\citenamefont {Regan},
  \citenamefont {Wang}, \citenamefont {Jin}, \citenamefont {Bakti~Utama},
  \citenamefont {Gao}, \citenamefont {Wei}, \citenamefont {Zhao}, \citenamefont
  {Zhao}, \citenamefont {Zhang}, \citenamefont {Yumigeta}, \citenamefont
  {Blei}, \citenamefont {Carlstr{\"o}m}, \citenamefont {Watanabe},
  \citenamefont {Taniguchi}, \citenamefont {Tongay}, \citenamefont {Crommie},
  \citenamefont {Zettl},\ and\ \citenamefont {Wang}}]{regan2020}%
  \BibitemOpen
  \bibfield  {author} {\bibinfo {author} {\bibfnamefont {E.~C.}\ \bibnamefont
  {Regan}}, \bibinfo {author} {\bibfnamefont {D.}~\bibnamefont {Wang}},
  \bibinfo {author} {\bibfnamefont {C.}~\bibnamefont {Jin}}, \bibinfo {author}
  {\bibfnamefont {M.~I.}\ \bibnamefont {Bakti~Utama}}, \bibinfo {author}
  {\bibfnamefont {B.}~\bibnamefont {Gao}}, \bibinfo {author} {\bibfnamefont
  {X.}~\bibnamefont {Wei}}, \bibinfo {author} {\bibfnamefont {S.}~\bibnamefont
  {Zhao}}, \bibinfo {author} {\bibfnamefont {W.}~\bibnamefont {Zhao}}, \bibinfo
  {author} {\bibfnamefont {Z.}~\bibnamefont {Zhang}}, \bibinfo {author}
  {\bibfnamefont {K.}~\bibnamefont {Yumigeta}}, \bibinfo {author}
  {\bibfnamefont {M.}~\bibnamefont {Blei}}, \bibinfo {author} {\bibfnamefont
  {J.~D.}\ \bibnamefont {Carlstr{\"o}m}}, \bibinfo {author} {\bibfnamefont
  {K.}~\bibnamefont {Watanabe}}, \bibinfo {author} {\bibfnamefont
  {T.}~\bibnamefont {Taniguchi}}, \bibinfo {author} {\bibfnamefont
  {S.}~\bibnamefont {Tongay}}, \bibinfo {author} {\bibfnamefont
  {M.}~\bibnamefont {Crommie}}, \bibinfo {author} {\bibfnamefont
  {A.}~\bibnamefont {Zettl}},\ and\ \bibinfo {author} {\bibfnamefont
  {F.}~\bibnamefont {Wang}},\ }\bibfield  {title} {\bibinfo {title} {Mott and
  generalized {{Wigner}} crystal states in {{WSe2}}/{{WS2}} moir{\'e}
  superlattices},\ }\href {https://doi.org/10.1038/s41586-020-2092-4}
  {\bibfield  {journal} {\bibinfo  {journal} {Nature}\ }\textbf {\bibinfo
  {volume} {579}},\ \bibinfo {pages} {359} (\bibinfo {year}
  {2020})}\BibitemShut {NoStop}%
\bibitem [{\citenamefont {Grimes}\ and\ \citenamefont
  {Adams}(1979)}]{grimes1979}%
  \BibitemOpen
  \bibfield  {author} {\bibinfo {author} {\bibfnamefont {C.~C.}\ \bibnamefont
  {Grimes}}\ and\ \bibinfo {author} {\bibfnamefont {G.}~\bibnamefont {Adams}},\
  }\bibfield  {title} {\bibinfo {title} {Evidence for a {{Liquid-to-Crystal
  Phase Transition}} in a {{Classical}}, {{Two-Dimensional Sheet}} of
  {{Electrons}}},\ }\href {https://doi.org/10.1103/PhysRevLett.42.795}
  {\bibfield  {journal} {\bibinfo  {journal} {Phys. Rev. Lett.}\ }\textbf
  {\bibinfo {volume} {42}},\ \bibinfo {pages} {795} (\bibinfo {year}
  {1979})}\BibitemShut {NoStop}%
\bibitem [{\citenamefont {Xu}\ \emph {et~al.}(2020)\citenamefont {Xu},
  \citenamefont {Liu}, \citenamefont {Rhodes}, \citenamefont {Watanabe},
  \citenamefont {Taniguchi}, \citenamefont {Hone}, \citenamefont {Elser},
  \citenamefont {Mak},\ and\ \citenamefont {Shan}}]{xu2020}%
  \BibitemOpen
  \bibfield  {author} {\bibinfo {author} {\bibfnamefont {Y.}~\bibnamefont
  {Xu}}, \bibinfo {author} {\bibfnamefont {S.}~\bibnamefont {Liu}}, \bibinfo
  {author} {\bibfnamefont {D.~A.}\ \bibnamefont {Rhodes}}, \bibinfo {author}
  {\bibfnamefont {K.}~\bibnamefont {Watanabe}}, \bibinfo {author}
  {\bibfnamefont {T.}~\bibnamefont {Taniguchi}}, \bibinfo {author}
  {\bibfnamefont {J.}~\bibnamefont {Hone}}, \bibinfo {author} {\bibfnamefont
  {V.}~\bibnamefont {Elser}}, \bibinfo {author} {\bibfnamefont {K.~F.}\
  \bibnamefont {Mak}},\ and\ \bibinfo {author} {\bibfnamefont {J.}~\bibnamefont
  {Shan}},\ }\bibfield  {title} {\bibinfo {title} {Correlated insulating states
  at fractional fillings of moir{\'e} superlattices},\ }\href
  {https://doi.org/10.1038/s41586-020-2868-6} {\bibfield  {journal} {\bibinfo
  {journal} {Nature}\ }\textbf {\bibinfo {volume} {587}},\ \bibinfo {pages}
  {214} (\bibinfo {year} {2020})}\BibitemShut {NoStop}%
\bibitem [{\citenamefont {Drewsen}(2015)}]{drewsen2015}%
  \BibitemOpen
  \bibfield  {author} {\bibinfo {author} {\bibfnamefont {M.}~\bibnamefont
  {Drewsen}},\ }\bibfield  {title} {\bibinfo {title} {Ion {{Coulomb}}
  crystals},\ }\href {https://doi.org/10.1016/j.physb.2014.11.050} {\bibfield
  {journal} {\bibinfo  {journal} {Physica B}\ }\textbf {\bibinfo {volume}
  {460}},\ \bibinfo {pages} {105} (\bibinfo {year} {2015})}\BibitemShut
  {NoStop}%
\bibitem [{\citenamefont {Schm{\"o}ger}\ \emph {et~al.}(2015)\citenamefont
  {Schm{\"o}ger}, \citenamefont {Versolato}, \citenamefont {Schwarz},
  \citenamefont {Kohnen}, \citenamefont {Windberger}, \citenamefont {Piest},
  \citenamefont {Feuchtenbeiner}, \citenamefont {{Pedregosa-Gutierrez}},
  \citenamefont {Leopold}, \citenamefont {Micke}, \citenamefont {Hansen},
  \citenamefont {Baumann}, \citenamefont {Drewsen}, \citenamefont {Ullrich},
  \citenamefont {Schmidt},\ and\ \citenamefont
  {{L{\'o}pez-Urrutia}}}]{schmoger2015}%
  \BibitemOpen
  \bibfield  {author} {\bibinfo {author} {\bibfnamefont {L.}~\bibnamefont
  {Schm{\"o}ger}}, \bibinfo {author} {\bibfnamefont {O.~O.}\ \bibnamefont
  {Versolato}}, \bibinfo {author} {\bibfnamefont {M.}~\bibnamefont {Schwarz}},
  \bibinfo {author} {\bibfnamefont {M.}~\bibnamefont {Kohnen}}, \bibinfo
  {author} {\bibfnamefont {A.}~\bibnamefont {Windberger}}, \bibinfo {author}
  {\bibfnamefont {B.}~\bibnamefont {Piest}}, \bibinfo {author} {\bibfnamefont
  {S.}~\bibnamefont {Feuchtenbeiner}}, \bibinfo {author} {\bibfnamefont
  {J.}~\bibnamefont {{Pedregosa-Gutierrez}}}, \bibinfo {author} {\bibfnamefont
  {T.}~\bibnamefont {Leopold}}, \bibinfo {author} {\bibfnamefont
  {P.}~\bibnamefont {Micke}}, \bibinfo {author} {\bibfnamefont {A.~K.}\
  \bibnamefont {Hansen}}, \bibinfo {author} {\bibfnamefont {T.~M.}\
  \bibnamefont {Baumann}}, \bibinfo {author} {\bibfnamefont {M.}~\bibnamefont
  {Drewsen}}, \bibinfo {author} {\bibfnamefont {J.}~\bibnamefont {Ullrich}},
  \bibinfo {author} {\bibfnamefont {P.~O.}\ \bibnamefont {Schmidt}},\ and\
  \bibinfo {author} {\bibfnamefont {J.~R.~C.}\ \bibnamefont
  {{L{\'o}pez-Urrutia}}},\ }\bibfield  {title} {\bibinfo {title} {Coulomb
  crystallization of highly charged ions},\ }\href
  {https://doi.org/10.1126/science.aaa2960} {\bibfield  {journal} {\bibinfo
  {journal} {Science}\ }\textbf {\bibinfo {volume} {347}},\ \bibinfo {pages}
  {1233} (\bibinfo {year} {2015})}\BibitemShut {NoStop}%
\bibitem [{\citenamefont {Drewsen}\ \emph {et~al.}(2003)\citenamefont
  {Drewsen}, \citenamefont {Jensen}, \citenamefont {Lindballe}, \citenamefont
  {Nissen}, \citenamefont {Martinussen}, \citenamefont {Mortensen},
  \citenamefont {Staanum},\ and\ \citenamefont {Voigt}}]{drewsen2003}%
  \BibitemOpen
  \bibfield  {author} {\bibinfo {author} {\bibfnamefont {M.}~\bibnamefont
  {Drewsen}}, \bibinfo {author} {\bibfnamefont {I.}~\bibnamefont {Jensen}},
  \bibinfo {author} {\bibfnamefont {J.}~\bibnamefont {Lindballe}}, \bibinfo
  {author} {\bibfnamefont {N.}~\bibnamefont {Nissen}}, \bibinfo {author}
  {\bibfnamefont {R.}~\bibnamefont {Martinussen}}, \bibinfo {author}
  {\bibfnamefont {A.}~\bibnamefont {Mortensen}}, \bibinfo {author}
  {\bibfnamefont {P.}~\bibnamefont {Staanum}},\ and\ \bibinfo {author}
  {\bibfnamefont {D.}~\bibnamefont {Voigt}},\ }\bibfield  {title} {\bibinfo
  {title} {Ion {{Coulomb}} crystals: A tool for studying ion processes},\
  }\href {https://doi.org/10.1016/S1387-3806(03)00259-8} {\bibfield  {journal}
  {\bibinfo  {journal} {Int. J. Mass Spectrom.}\ }\textbf {\bibinfo {volume}
  {229}},\ \bibinfo {pages} {83} (\bibinfo {year} {2003})}\BibitemShut
  {NoStop}%
\bibitem [{\citenamefont {Li}\ \emph {et~al.}(2021)\citenamefont {Li},
  \citenamefont {Li}, \citenamefont {Regan}, \citenamefont {Wang},
  \citenamefont {Zhao}, \citenamefont {Kahn}, \citenamefont {Yumigeta},
  \citenamefont {Blei}, \citenamefont {Taniguchi}, \citenamefont {Watanabe},
  \citenamefont {Tongay}, \citenamefont {Zettl}, \citenamefont {Crommie},\ and\
  \citenamefont {Wang}}]{li2021}%
  \BibitemOpen
  \bibfield  {author} {\bibinfo {author} {\bibfnamefont {H.}~\bibnamefont
  {Li}}, \bibinfo {author} {\bibfnamefont {S.}~\bibnamefont {Li}}, \bibinfo
  {author} {\bibfnamefont {E.~C.}\ \bibnamefont {Regan}}, \bibinfo {author}
  {\bibfnamefont {D.}~\bibnamefont {Wang}}, \bibinfo {author} {\bibfnamefont
  {W.}~\bibnamefont {Zhao}}, \bibinfo {author} {\bibfnamefont {S.}~\bibnamefont
  {Kahn}}, \bibinfo {author} {\bibfnamefont {K.}~\bibnamefont {Yumigeta}},
  \bibinfo {author} {\bibfnamefont {M.}~\bibnamefont {Blei}}, \bibinfo {author}
  {\bibfnamefont {T.}~\bibnamefont {Taniguchi}}, \bibinfo {author}
  {\bibfnamefont {K.}~\bibnamefont {Watanabe}}, \bibinfo {author}
  {\bibfnamefont {S.}~\bibnamefont {Tongay}}, \bibinfo {author} {\bibfnamefont
  {A.}~\bibnamefont {Zettl}}, \bibinfo {author} {\bibfnamefont {M.~F.}\
  \bibnamefont {Crommie}},\ and\ \bibinfo {author} {\bibfnamefont
  {F.}~\bibnamefont {Wang}},\ }\bibfield  {title} {\bibinfo {title} {Imaging
  two-dimensional generalized {{Wigner}} crystals},\ }\href
  {https://doi.org/10.1038/s41586-021-03874-9} {\bibfield  {journal} {\bibinfo
  {journal} {Nature}\ }\textbf {\bibinfo {volume} {597}},\ \bibinfo {pages}
  {650} (\bibinfo {year} {2021})}\BibitemShut {NoStop}%
\bibitem [{\citenamefont {Baus}(1980)}]{baus1980}%
  \BibitemOpen
  \bibfield  {author} {\bibinfo {author} {\bibfnamefont {M.}~\bibnamefont
  {Baus}},\ }\bibfield  {title} {\bibinfo {title} {Statistical mechanics of
  simple coulomb systems},\ }\href
  {https://doi.org/10.1016/0370-1573(80)90022-8} {\bibfield  {journal}
  {\bibinfo  {journal} {Phys. Rep.}\ }\textbf {\bibinfo {volume} {59}},\
  \bibinfo {pages} {1} (\bibinfo {year} {1980})}\BibitemShut {NoStop}%
\bibitem [{\citenamefont {Saigo}\ and\ \citenamefont
  {Hamaguchi}(2002)}]{saigo2002}%
  \BibitemOpen
  \bibfield  {author} {\bibinfo {author} {\bibfnamefont {T.}~\bibnamefont
  {Saigo}}\ and\ \bibinfo {author} {\bibfnamefont {S.}~\bibnamefont
  {Hamaguchi}},\ }\bibfield  {title} {\bibinfo {title} {Shear viscosity of
  strongly coupled {{Yukawa}} systems},\ }\href
  {https://doi.org/10.1063/1.1459708} {\bibfield  {journal} {\bibinfo
  {journal} {Phys. Plasmas}\ }\textbf {\bibinfo {volume} {9}},\ \bibinfo
  {pages} {1210} (\bibinfo {year} {2002})}\BibitemShut {NoStop}%
\bibitem [{\citenamefont {Vieillefosse}\ and\ \citenamefont
  {Hansen}(1975)}]{vieillefosse1975}%
  \BibitemOpen
  \bibfield  {author} {\bibinfo {author} {\bibfnamefont {P.}~\bibnamefont
  {Vieillefosse}}\ and\ \bibinfo {author} {\bibfnamefont {J.~P.}\ \bibnamefont
  {Hansen}},\ }\bibfield  {title} {\bibinfo {title} {Statistical mechanics of
  dense ionized matter. {{V}}. {{Hydrodynamic}} limit and transport
  coefficients of the classical one-component plasma},\ }\href
  {https://doi.org/10.1103/PhysRevA.12.1106} {\bibfield  {journal} {\bibinfo
  {journal} {Phys. Rev. A}\ }\textbf {\bibinfo {volume} {12}},\ \bibinfo
  {pages} {1106} (\bibinfo {year} {1975})}\BibitemShut {NoStop}%
\bibitem [{\citenamefont {Hansen}(1973)}]{hansen1973}%
  \BibitemOpen
  \bibfield  {author} {\bibinfo {author} {\bibfnamefont {J.~P.}\ \bibnamefont
  {Hansen}},\ }\bibfield  {title} {\bibinfo {title} {Statistical {{Mechanics}}
  of {{Dense Ionized Matter}}. {{I}}. {{Equilibrium Properties}} of the
  {{Classical One-Component Plasma}}},\ }\href
  {https://doi.org/10.1103/PhysRevA.8.3096} {\bibfield  {journal} {\bibinfo
  {journal} {Phys. Rev. A}\ }\textbf {\bibinfo {volume} {8}},\ \bibinfo {pages}
  {3096} (\bibinfo {year} {1973})}\BibitemShut {NoStop}%
\bibitem [{\citenamefont {Pollock}\ and\ \citenamefont
  {Hansen}(1973)}]{pollock1973}%
  \BibitemOpen
  \bibfield  {author} {\bibinfo {author} {\bibfnamefont {E.~L.}\ \bibnamefont
  {Pollock}}\ and\ \bibinfo {author} {\bibfnamefont {J.~P.}\ \bibnamefont
  {Hansen}},\ }\bibfield  {title} {\bibinfo {title} {Statistical {{Mechanics}}
  of {{Dense Ionized Matter}}. {{II}}. {{Equilibrium Properties}} and {{Melting
  Transition}} of the {{Crystallized One-Component Plasma}}},\ }\href
  {https://doi.org/10.1103/PhysRevA.8.3110} {\bibfield  {journal} {\bibinfo
  {journal} {Phys. Rev. A}\ }\textbf {\bibinfo {volume} {8}},\ \bibinfo {pages}
  {3110} (\bibinfo {year} {1973})}\BibitemShut {NoStop}%
\bibitem [{\citenamefont {Ichimaru}(1982)}]{ichimaru1982}%
  \BibitemOpen
  \bibfield  {author} {\bibinfo {author} {\bibfnamefont {S.}~\bibnamefont
  {Ichimaru}},\ }\bibfield  {title} {\bibinfo {title} {Strongly coupled
  plasmas: High-density classical plasmas and degenerate electron liquids},\
  }\href {https://doi.org/10.1103/RevModPhys.54.1017} {\bibfield  {journal}
  {\bibinfo  {journal} {Rev. Mod. Phys.}\ }\textbf {\bibinfo {volume} {54}},\
  \bibinfo {pages} {1017} (\bibinfo {year} {1982})}\BibitemShut {NoStop}%
\bibitem [{\citenamefont {Schiffer}(2003)}]{schiffer2003}%
  \BibitemOpen
  \bibfield  {author} {\bibinfo {author} {\bibfnamefont {J.~P.}\ \bibnamefont
  {Schiffer}},\ }\bibfield  {title} {\bibinfo {title} {Order in confined
  ions},\ }\href {https://doi.org/10.1088/0953-4075/36/3/309} {\bibfield
  {journal} {\bibinfo  {journal} {J. Phys. B: At. Mol. Opt. Phys.}\ }\textbf
  {\bibinfo {volume} {36}},\ \bibinfo {pages} {511} (\bibinfo {year}
  {2003})}\BibitemShut {NoStop}%
\bibitem [{\citenamefont {Rogers}\ and\ \citenamefont
  {Dewitt}(1987)}]{rogers1987}%
  \BibitemOpen
  \bibinfo {editor} {\bibfnamefont {F.~J.}\ \bibnamefont {Rogers}}\ and\
  \bibinfo {editor} {\bibfnamefont {H.~E.}\ \bibnamefont {Dewitt}},\ eds.,\
  \href {https://doi.org/10.1007/978-1-4613-1891-0} {\emph {\bibinfo {title}
  {Strongly {{Coupled Plasma Physics}}}}}\ (\bibinfo  {publisher} {Springer
  US},\ \bibinfo {address} {Boston, MA},\ \bibinfo {year} {1987})\BibitemShut
  {NoStop}%
\bibitem [{\citenamefont {Bernu}\ and\ \citenamefont
  {Vieillefosse}(1978)}]{bernu1978}%
  \BibitemOpen
  \bibfield  {author} {\bibinfo {author} {\bibfnamefont {B.}~\bibnamefont
  {Bernu}}\ and\ \bibinfo {author} {\bibfnamefont {P.}~\bibnamefont
  {Vieillefosse}},\ }\bibfield  {title} {\bibinfo {title} {Transport
  coefficients of the classical one-component plasma},\ }\href
  {https://doi.org/10.1103/PhysRevA.18.2345} {\bibfield  {journal} {\bibinfo
  {journal} {Phys. Rev. A}\ }\textbf {\bibinfo {volume} {18}},\ \bibinfo
  {pages} {2345} (\bibinfo {year} {1978})}\BibitemShut {NoStop}%
\bibitem [{\citenamefont {Donk{\'o}}\ \emph {et~al.}(1998)\citenamefont
  {Donk{\'o}}, \citenamefont {Ny{\'i}ri}, \citenamefont {Szalai},\ and\
  \citenamefont {Holl{\'o}}}]{donko1998}%
  \BibitemOpen
  \bibfield  {author} {\bibinfo {author} {\bibfnamefont {Z.}~\bibnamefont
  {Donk{\'o}}}, \bibinfo {author} {\bibfnamefont {B.}~\bibnamefont
  {Ny{\'i}ri}}, \bibinfo {author} {\bibfnamefont {L.}~\bibnamefont {Szalai}},\
  and\ \bibinfo {author} {\bibfnamefont {S.}~\bibnamefont {Holl{\'o}}},\
  }\bibfield  {title} {\bibinfo {title} {Thermal {{Conductivity}} of the
  {{Classical Electron One-Component Plasma}}},\ }\href
  {https://doi.org/10.1103/PhysRevLett.81.1622} {\bibfield  {journal} {\bibinfo
   {journal} {Phys. Rev. Lett.}\ }\textbf {\bibinfo {volume} {81}},\ \bibinfo
  {pages} {1622} (\bibinfo {year} {1998})}\BibitemShut {NoStop}%
\bibitem [{\citenamefont {Donk{\'o}}\ and\ \citenamefont
  {Hartmann}(2004)}]{donko2004}%
  \BibitemOpen
  \bibfield  {author} {\bibinfo {author} {\bibfnamefont {Z.}~\bibnamefont
  {Donk{\'o}}}\ and\ \bibinfo {author} {\bibfnamefont {P.}~\bibnamefont
  {Hartmann}},\ }\bibfield  {title} {\bibinfo {title} {Thermal conductivity of
  strongly coupled {{Yukawa}} liquids},\ }\href
  {https://doi.org/10.1103/PhysRevE.69.016405} {\bibfield  {journal} {\bibinfo
  {journal} {Phys. Rev. E}\ }\textbf {\bibinfo {volume} {69}},\ \bibinfo
  {pages} {016405} (\bibinfo {year} {2004})}\BibitemShut {NoStop}%
\bibitem [{\citenamefont {Daligault}\ \emph {et~al.}(2014)\citenamefont
  {Daligault}, \citenamefont {Rasmussen},\ and\ \citenamefont
  {Baalrud}}]{daligault2014}%
  \BibitemOpen
  \bibfield  {author} {\bibinfo {author} {\bibfnamefont {J.}~\bibnamefont
  {Daligault}}, \bibinfo {author} {\bibfnamefont {K.~{\O}.}\ \bibnamefont
  {Rasmussen}},\ and\ \bibinfo {author} {\bibfnamefont {S.~D.}\ \bibnamefont
  {Baalrud}},\ }\bibfield  {title} {\bibinfo {title} {Determination of the
  shear viscosity of the one-component plasma},\ }\href
  {https://doi.org/10.1103/PhysRevE.90.033105} {\bibfield  {journal} {\bibinfo
  {journal} {Phys. Rev. E}\ }\textbf {\bibinfo {volume} {90}},\ \bibinfo
  {pages} {033105} (\bibinfo {year} {2014})}\BibitemShut {NoStop}%
\bibitem [{\citenamefont {Kamimura}\ and\ \citenamefont
  {Ishihara}(2012)}]{kamimura2012}%
  \BibitemOpen
  \bibfield  {author} {\bibinfo {author} {\bibfnamefont {T.}~\bibnamefont
  {Kamimura}}\ and\ \bibinfo {author} {\bibfnamefont {O.}~\bibnamefont
  {Ishihara}},\ }\bibfield  {title} {\bibinfo {title} {Coulomb double helical
  structure},\ }\href {https://doi.org/10.1103/PhysRevE.85.016406} {\bibfield
  {journal} {\bibinfo  {journal} {Phys. Rev. E}\ }\textbf {\bibinfo {volume}
  {85}},\ \bibinfo {pages} {016406} (\bibinfo {year} {2012})}\BibitemShut
  {NoStop}%
\bibitem [{\citenamefont {Hyde}\ \emph {et~al.}(2013)\citenamefont {Hyde},
  \citenamefont {Kong},\ and\ \citenamefont {Matthews}}]{hyde2013}%
  \BibitemOpen
  \bibfield  {author} {\bibinfo {author} {\bibfnamefont {T.~W.}\ \bibnamefont
  {Hyde}}, \bibinfo {author} {\bibfnamefont {J.}~\bibnamefont {Kong}},\ and\
  \bibinfo {author} {\bibfnamefont {L.~S.}\ \bibnamefont {Matthews}},\
  }\bibfield  {title} {\bibinfo {title} {Helical structures in vertically
  aligned dust particle chains in a complex plasma},\ }\href
  {https://doi.org/10.1103/PhysRevE.87.053106} {\bibfield  {journal} {\bibinfo
  {journal} {Phys. Rev. E}\ }\textbf {\bibinfo {volume} {87}},\ \bibinfo
  {pages} {053106} (\bibinfo {year} {2013})}\BibitemShut {NoStop}%
\bibitem [{\citenamefont {{Gusein-zade}}\ and\ \citenamefont
  {Tsytovich}(2006)}]{gusein-zade2006}%
  \BibitemOpen
  \bibfield  {author} {\bibinfo {author} {\bibfnamefont {N.~G.}\ \bibnamefont
  {{Gusein-zade}}}\ and\ \bibinfo {author} {\bibfnamefont {V.~N.}\ \bibnamefont
  {Tsytovich}},\ }\bibfield  {title} {\bibinfo {title} {Spectral properties of
  an {{N-pole}} helical structure consisting of like-charged equal-size dust
  grains},\ }\href {https://doi.org/10.1134/S1063780X06080046} {\bibfield
  {journal} {\bibinfo  {journal} {Plasma Phys. Rep.}\ }\textbf {\bibinfo
  {volume} {32}},\ \bibinfo {pages} {668} (\bibinfo {year} {2006})}\BibitemShut
  {NoStop}%
\bibitem [{\citenamefont {Landa}\ \emph {et~al.}(2013)\citenamefont {Landa},
  \citenamefont {Reznik}, \citenamefont {Brox}, \citenamefont {Mielenz},\ and\
  \citenamefont {Schaetz}}]{landa2013}%
  \BibitemOpen
  \bibfield  {author} {\bibinfo {author} {\bibfnamefont {H.}~\bibnamefont
  {Landa}}, \bibinfo {author} {\bibfnamefont {B.}~\bibnamefont {Reznik}},
  \bibinfo {author} {\bibfnamefont {J.}~\bibnamefont {Brox}}, \bibinfo {author}
  {\bibfnamefont {M.}~\bibnamefont {Mielenz}},\ and\ \bibinfo {author}
  {\bibfnamefont {T.}~\bibnamefont {Schaetz}},\ }\bibfield  {title} {\bibinfo
  {title} {Structure, dynamics and bifurcations of discrete solitons in trapped
  ion crystals},\ }\href {https://doi.org/10.1088/1367-2630/15/9/093003}
  {\bibfield  {journal} {\bibinfo  {journal} {New J. Phys.}\ }\textbf {\bibinfo
  {volume} {15}},\ \bibinfo {pages} {093003} (\bibinfo {year}
  {2013})}\BibitemShut {NoStop}%
\bibitem [{\citenamefont {D'yachkov}\ \emph {et~al.}(2014)\citenamefont
  {D'yachkov}, \citenamefont {Myasnikov}, \citenamefont {Petrov}, \citenamefont
  {Hyde}, \citenamefont {Kong},\ and\ \citenamefont {Matthews}}]{dyachkov2014}%
  \BibitemOpen
  \bibfield  {author} {\bibinfo {author} {\bibfnamefont {L.~G.}\ \bibnamefont
  {D'yachkov}}, \bibinfo {author} {\bibfnamefont {M.~I.}\ \bibnamefont
  {Myasnikov}}, \bibinfo {author} {\bibfnamefont {O.~F.}\ \bibnamefont
  {Petrov}}, \bibinfo {author} {\bibfnamefont {T.~W.}\ \bibnamefont {Hyde}},
  \bibinfo {author} {\bibfnamefont {J.}~\bibnamefont {Kong}},\ and\ \bibinfo
  {author} {\bibfnamefont {L.}~\bibnamefont {Matthews}},\ }\bibfield  {title}
  {\bibinfo {title} {Two-dimensional and three-dimensional {{Coulomb}} clusters
  in parabolic traps},\ }\href {https://doi.org/10.1063/1.4885637} {\bibfield
  {journal} {\bibinfo  {journal} {Phys. Plasmas}\ }\textbf {\bibinfo {volume}
  {21}},\ \bibinfo {pages} {093702} (\bibinfo {year} {2014})}\BibitemShut
  {NoStop}%
\bibitem [{\citenamefont {Piacente}\ \emph {et~al.}(2004)\citenamefont
  {Piacente}, \citenamefont {Schweigert}, \citenamefont {Betouras},\ and\
  \citenamefont {Peeters}}]{piacente2004}%
  \BibitemOpen
  \bibfield  {author} {\bibinfo {author} {\bibfnamefont {G.}~\bibnamefont
  {Piacente}}, \bibinfo {author} {\bibfnamefont {I.~V.}\ \bibnamefont
  {Schweigert}}, \bibinfo {author} {\bibfnamefont {J.~J.}\ \bibnamefont
  {Betouras}},\ and\ \bibinfo {author} {\bibfnamefont {F.~M.}\ \bibnamefont
  {Peeters}},\ }\bibfield  {title} {\bibinfo {title} {Generic properties of a
  quasi-one-dimensional classical {{Wigner}} crystal},\ }\href
  {https://doi.org/10.1103/PhysRevB.69.045324} {\bibfield  {journal} {\bibinfo
  {journal} {Phys. Rev. B}\ }\textbf {\bibinfo {volume} {69}},\ \bibinfo
  {pages} {045324} (\bibinfo {year} {2004})}\BibitemShut {NoStop}%
\bibitem [{\citenamefont {Birkl}\ \emph {et~al.}(1992)\citenamefont {Birkl},
  \citenamefont {Kassner},\ and\ \citenamefont {Walther}}]{birkl1992}%
  \BibitemOpen
  \bibfield  {author} {\bibinfo {author} {\bibfnamefont {G.}~\bibnamefont
  {Birkl}}, \bibinfo {author} {\bibfnamefont {S.}~\bibnamefont {Kassner}},\
  and\ \bibinfo {author} {\bibfnamefont {H.}~\bibnamefont {Walther}},\
  }\bibfield  {title} {\bibinfo {title} {Multiple-shell structures of
  laser-cooled {{24Mg}}+ ions in a quadrupole storage ring},\ }\href
  {https://doi.org/10.1038/357310a0} {\bibfield  {journal} {\bibinfo  {journal}
  {Nature}\ }\textbf {\bibinfo {volume} {357}},\ \bibinfo {pages} {310}
  (\bibinfo {year} {1992})}\BibitemShut {NoStop}%
\bibitem [{\citenamefont {Bollinger}\ \emph {et~al.}(1994)\citenamefont
  {Bollinger}, \citenamefont {Wineland},\ and\ \citenamefont
  {Dubin}}]{bollinger1994}%
  \BibitemOpen
  \bibfield  {author} {\bibinfo {author} {\bibfnamefont {J.~J.}\ \bibnamefont
  {Bollinger}}, \bibinfo {author} {\bibfnamefont {D.~J.}\ \bibnamefont
  {Wineland}},\ and\ \bibinfo {author} {\bibfnamefont {D.~H.~E.}\ \bibnamefont
  {Dubin}},\ }\bibfield  {title} {\bibinfo {title} {Non-neutral ion plasmas and
  crystals, laser cooling, and atomic clocks*},\ }\href
  {https://doi.org/10.1063/1.870690} {\bibfield  {journal} {\bibinfo  {journal}
  {Phys. Plasmas}\ }\textbf {\bibinfo {volume} {1}},\ \bibinfo {pages} {1403}
  (\bibinfo {year} {1994})}\BibitemShut {NoStop}%
\bibitem [{\citenamefont {Hasse}\ and\ \citenamefont
  {Schiffer}(1990)}]{hasse1990}%
  \BibitemOpen
  \bibfield  {author} {\bibinfo {author} {\bibfnamefont {R.}~\bibnamefont
  {Hasse}}\ and\ \bibinfo {author} {\bibfnamefont {J.}~\bibnamefont
  {Schiffer}},\ }\bibfield  {title} {\bibinfo {title} {The structure of the
  cylindrically confined {{Coulomb}} lattice},\ }\href
  {https://doi.org/10.1016/0003-4916(90)90177-P} {\bibfield  {journal}
  {\bibinfo  {journal} {Ann. Phys.}\ }\textbf {\bibinfo {volume} {203}},\
  \bibinfo {pages} {419} (\bibinfo {year} {1990})}\BibitemShut {NoStop}%
\bibitem [{\citenamefont {{Gusein-zade}}(2005)}]{gusein-zade2005}%
  \BibitemOpen
  \bibfield  {author} {\bibinfo {author} {\bibfnamefont {N.~G.}\ \bibnamefont
  {{Gusein-zade}}},\ }\bibfield  {title} {\bibinfo {title} {Helical
  {{Structures}} in {{Complex Plasma I}}: {{Noncollective Interaction}}},\
  }\href {https://doi.org/10.1134/1.1925789} {\bibfield  {journal} {\bibinfo
  {journal} {Plasma Phys. Rep.}\ }\textbf {\bibinfo {volume} {31}},\ \bibinfo
  {pages} {392} (\bibinfo {year} {2005})}\BibitemShut {NoStop}%
\bibitem [{\citenamefont {Tsytovich}(2005)}]{tsytovich2005}%
  \BibitemOpen
  \bibfield  {author} {\bibinfo {author} {\bibfnamefont {V.~N.}\ \bibnamefont
  {Tsytovich}},\ }\bibfield  {title} {\bibinfo {title} {Helical {{Structures}}
  in {{Complex Plasma II}}: {{Collective Interaction}}},\ }\href
  {https://doi.org/10.1134/1.2101970} {\bibfield  {journal} {\bibinfo
  {journal} {Plasma Phys. Rep.}\ }\textbf {\bibinfo {volume} {31}},\ \bibinfo
  {pages} {824} (\bibinfo {year} {2005})}\BibitemShut {NoStop}%
\bibitem [{\citenamefont {Fortov}\ \emph {et~al.}(2004)\citenamefont {Fortov},
  \citenamefont {Khrapak}, \citenamefont {Khrapak}, \citenamefont {Molotkov},\
  and\ \citenamefont {Petrov}}]{fortov2004}%
  \BibitemOpen
  \bibfield  {author} {\bibinfo {author} {\bibfnamefont {V.~E.}\ \bibnamefont
  {Fortov}}, \bibinfo {author} {\bibfnamefont {A.~G.}\ \bibnamefont {Khrapak}},
  \bibinfo {author} {\bibfnamefont {S.~A.}\ \bibnamefont {Khrapak}}, \bibinfo
  {author} {\bibfnamefont {V.~I.}\ \bibnamefont {Molotkov}},\ and\ \bibinfo
  {author} {\bibfnamefont {O.~F.}\ \bibnamefont {Petrov}},\ }\bibfield  {title}
  {\bibinfo {title} {Dusty plasmas},\ }\href
  {https://doi.org/10.1070/PU2004v047n05ABEH001689} {\bibfield  {journal}
  {\bibinfo  {journal} {Phys.-Usp.}\ }\textbf {\bibinfo {volume} {47}},\
  \bibinfo {pages} {447} (\bibinfo {year} {2004})}\BibitemShut {NoStop}%
\bibitem [{\citenamefont {Fortov}\ \emph {et~al.}(2005)\citenamefont {Fortov},
  \citenamefont {Ivlev}, \citenamefont {Khrapak}, \citenamefont {Khrapak},\
  and\ \citenamefont {Morfill}}]{fortov2005}%
  \BibitemOpen
  \bibfield  {author} {\bibinfo {author} {\bibfnamefont {V.}~\bibnamefont
  {Fortov}}, \bibinfo {author} {\bibfnamefont {A.}~\bibnamefont {Ivlev}},
  \bibinfo {author} {\bibfnamefont {S.}~\bibnamefont {Khrapak}}, \bibinfo
  {author} {\bibfnamefont {A.}~\bibnamefont {Khrapak}},\ and\ \bibinfo {author}
  {\bibfnamefont {G.}~\bibnamefont {Morfill}},\ }\bibfield  {title} {\bibinfo
  {title} {Complex (dusty) plasmas: {{Current}} status, open issues,
  perspectives},\ }\href {https://doi.org/10.1016/j.physrep.2005.08.007}
  {\bibfield  {journal} {\bibinfo  {journal} {Phys. Rep.}\ }\textbf {\bibinfo
  {volume} {421}},\ \bibinfo {pages} {1} (\bibinfo {year} {2005})}\BibitemShut
  {NoStop}%
\bibitem [{\citenamefont {Ignatov}(2005)}]{ignatov2005}%
  \BibitemOpen
  \bibfield  {author} {\bibinfo {author} {\bibfnamefont {A.~M.}\ \bibnamefont
  {Ignatov}},\ }\bibfield  {title} {\bibinfo {title} {Basics of dusty plasma},\
  }\href {https://doi.org/10.1134/1.1856707} {\bibfield  {journal} {\bibinfo
  {journal} {Plasma Phys. Rep.}\ }\textbf {\bibinfo {volume} {31}},\ \bibinfo
  {pages} {46} (\bibinfo {year} {2005})}\BibitemShut {NoStop}%
\bibitem [{\citenamefont {Schmelcher}(2011)}]{schmelcher2011}%
  \BibitemOpen
  \bibfield  {author} {\bibinfo {author} {\bibfnamefont {P.}~\bibnamefont
  {Schmelcher}},\ }\bibfield  {title} {\bibinfo {title} {Effective long-range
  interactions in confined curved dimensions},\ }\href
  {https://doi.org/10.1209/0295-5075/95/50005} {\bibfield  {journal} {\bibinfo
  {journal} {Europhys. Lett.}\ }\textbf {\bibinfo {volume} {95}},\ \bibinfo
  {pages} {50005} (\bibinfo {year} {2011})}\BibitemShut {NoStop}%
\bibitem [{\citenamefont {Kibis}(1992)}]{kibis1992}%
  \BibitemOpen
  \bibfield  {author} {\bibinfo {author} {\bibfnamefont {O.}~\bibnamefont
  {Kibis}},\ }\bibfield  {title} {\bibinfo {title} {Electron-electron
  interaction in a spiral quantum wire},\ }\href
  {https://doi.org/10.1016/0375-9601(92)90730-A} {\bibfield  {journal}
  {\bibinfo  {journal} {Phys. Lett. A}\ }\textbf {\bibinfo {volume} {166}},\
  \bibinfo {pages} {393} (\bibinfo {year} {1992})}\BibitemShut {NoStop}%
\bibitem [{\citenamefont {Tokihiro}\ and\ \citenamefont
  {Hanamura}(1993)}]{tokihiro1993}%
  \BibitemOpen
  \bibfield  {author} {\bibinfo {author} {\bibfnamefont {T.}~\bibnamefont
  {Tokihiro}}\ and\ \bibinfo {author} {\bibfnamefont {E.}~\bibnamefont
  {Hanamura}},\ }\bibfield  {title} {\bibinfo {title} {Geometrical effects on
  one-dimensional excitons},\ }\href
  {https://doi.org/10.1103/PhysRevLett.71.1423} {\bibfield  {journal} {\bibinfo
   {journal} {Phys. Rev. Lett.}\ }\textbf {\bibinfo {volume} {71}},\ \bibinfo
  {pages} {1423} (\bibinfo {year} {1993})}\BibitemShut {NoStop}%
\bibitem [{\citenamefont {Sarma}\ and\ \citenamefont
  {Gartstein}(2010)}]{sarma2010}%
  \BibitemOpen
  \bibfield  {author} {\bibinfo {author} {\bibfnamefont {R.}~\bibnamefont
  {Sarma}}\ and\ \bibinfo {author} {\bibfnamefont {{\relax Yu.N}.}~\bibnamefont
  {Gartstein}},\ }\bibfield  {title} {\bibinfo {title} {Wannier--{{Mott}}
  excitons on a helically-shaped one-dimensional semiconductor},\ }\href
  {https://doi.org/10.1016/j.physleta.2010.05.032} {\bibfield  {journal}
  {\bibinfo  {journal} {Phys. Lett. A}\ }\textbf {\bibinfo {volume} {374}},\
  \bibinfo {pages} {3076} (\bibinfo {year} {2010})}\BibitemShut {NoStop}%
\bibitem [{\citenamefont {Pedersen}\ \emph {et~al.}(2014)\citenamefont
  {Pedersen}, \citenamefont {Fedorov}, \citenamefont {Jensen},\ and\
  \citenamefont {Zinner}}]{pedersen2014}%
  \BibitemOpen
  \bibfield  {author} {\bibinfo {author} {\bibfnamefont {J.~K.}\ \bibnamefont
  {Pedersen}}, \bibinfo {author} {\bibfnamefont {D.~V.}\ \bibnamefont
  {Fedorov}}, \bibinfo {author} {\bibfnamefont {A.~S.}\ \bibnamefont
  {Jensen}},\ and\ \bibinfo {author} {\bibfnamefont {N.~T.}\ \bibnamefont
  {Zinner}},\ }\bibfield  {title} {\bibinfo {title} {Formation of classical
  crystals of dipolar particles in a helical geometry},\ }\href
  {https://doi.org/10.1088/0953-4075/47/16/165103} {\bibfield  {journal}
  {\bibinfo  {journal} {J. Phys. B: At. Mol. Opt. Phys.}\ }\textbf {\bibinfo
  {volume} {47}},\ \bibinfo {pages} {165103} (\bibinfo {year}
  {2014})}\BibitemShut {NoStop}%
\bibitem [{\citenamefont {Pedersen}\ \emph
  {et~al.}(2016{\natexlab{a}})\citenamefont {Pedersen}, \citenamefont
  {Fedorov}, \citenamefont {Jensen},\ and\ \citenamefont
  {Zinner}}]{pedersen2016}%
  \BibitemOpen
  \bibfield  {author} {\bibinfo {author} {\bibfnamefont {J.~K.}\ \bibnamefont
  {Pedersen}}, \bibinfo {author} {\bibfnamefont {D.~V.}\ \bibnamefont
  {Fedorov}}, \bibinfo {author} {\bibfnamefont {A.~S.}\ \bibnamefont
  {Jensen}},\ and\ \bibinfo {author} {\bibfnamefont {N.~T.}\ \bibnamefont
  {Zinner}},\ }\bibfield  {title} {\bibinfo {title} {Quantum few-body bound
  states of dipolar particles in a helical geometry},\ }\href
  {https://doi.org/10.1088/0953-4075/49/2/024002} {\bibfield  {journal}
  {\bibinfo  {journal} {J. Phys. B: At. Mol. Opt. Phys.}\ }\textbf {\bibinfo
  {volume} {49}},\ \bibinfo {pages} {024002} (\bibinfo {year}
  {2016}{\natexlab{a}})}\BibitemShut {NoStop}%
\bibitem [{\citenamefont {Pedersen}\ \emph
  {et~al.}(2016{\natexlab{b}})\citenamefont {Pedersen}, \citenamefont
  {Fedorov}, \citenamefont {Jensen},\ and\ \citenamefont
  {Zinner}}]{pedersen2016a}%
  \BibitemOpen
  \bibfield  {author} {\bibinfo {author} {\bibfnamefont {J.~K.}\ \bibnamefont
  {Pedersen}}, \bibinfo {author} {\bibfnamefont {D.~V.}\ \bibnamefont
  {Fedorov}}, \bibinfo {author} {\bibfnamefont {A.~S.}\ \bibnamefont
  {Jensen}},\ and\ \bibinfo {author} {\bibfnamefont {N.~T.}\ \bibnamefont
  {Zinner}},\ }\bibfield  {title} {\bibinfo {title} {Quantum single-particle
  properties in a one-dimensional curved space},\ }\href
  {https://doi.org/10.1080/09500340.2015.1116634} {\bibfield  {journal}
  {\bibinfo  {journal} {J. Mod. Opt.}\ }\textbf {\bibinfo {volume} {63}},\
  \bibinfo {pages} {1814} (\bibinfo {year} {2016}{\natexlab{b}})}\BibitemShut
  {NoStop}%
\bibitem [{\citenamefont {Law}\ and\ \citenamefont {Feldman}(2008)}]{law2008}%
  \BibitemOpen
  \bibfield  {author} {\bibinfo {author} {\bibfnamefont {K.~T.}\ \bibnamefont
  {Law}}\ and\ \bibinfo {author} {\bibfnamefont {D.~E.}\ \bibnamefont
  {Feldman}},\ }\bibfield  {title} {\bibinfo {title} {Quantum {{Phase
  Transition Between}} a {{Luttinger Liquid}} and a {{Gas}} of {{Cold
  Molecules}}},\ }\href {https://doi.org/10.1103/PhysRevLett.101.096401}
  {\bibfield  {journal} {\bibinfo  {journal} {Phys. Rev. Lett.}\ }\textbf
  {\bibinfo {volume} {101}},\ \bibinfo {pages} {096401} (\bibinfo {year}
  {2008})}\BibitemShut {NoStop}%
\bibitem [{\citenamefont {Zampetaki}\ \emph
  {et~al.}(2015{\natexlab{a}})\citenamefont {Zampetaki}, \citenamefont
  {Stockhofe},\ and\ \citenamefont {Schmelcher}}]{zampetaki2015}%
  \BibitemOpen
  \bibfield  {author} {\bibinfo {author} {\bibfnamefont {A.~V.}\ \bibnamefont
  {Zampetaki}}, \bibinfo {author} {\bibfnamefont {J.}~\bibnamefont
  {Stockhofe}},\ and\ \bibinfo {author} {\bibfnamefont {P.}~\bibnamefont
  {Schmelcher}},\ }\bibfield  {title} {\bibinfo {title} {Degeneracy and
  inversion of band structure for {{Wigner}} crystals on a closed helix},\
  }\href {https://doi.org/10.1103/PhysRevA.91.023409} {\bibfield  {journal}
  {\bibinfo  {journal} {Phys. Rev. A}\ }\textbf {\bibinfo {volume} {91}},\
  \bibinfo {pages} {023409} (\bibinfo {year} {2015}{\natexlab{a}})}\BibitemShut
  {NoStop}%
\bibitem [{\citenamefont {Zampetaki}\ \emph
  {et~al.}(2015{\natexlab{b}})\citenamefont {Zampetaki}, \citenamefont
  {Stockhofe},\ and\ \citenamefont {Schmelcher}}]{zampetaki2015a}%
  \BibitemOpen
  \bibfield  {author} {\bibinfo {author} {\bibfnamefont {A.~V.}\ \bibnamefont
  {Zampetaki}}, \bibinfo {author} {\bibfnamefont {J.}~\bibnamefont
  {Stockhofe}},\ and\ \bibinfo {author} {\bibfnamefont {P.}~\bibnamefont
  {Schmelcher}},\ }\bibfield  {title} {\bibinfo {title} {Dynamics of nonlinear
  excitations of helically confined charges},\ }\href
  {https://doi.org/10.1103/PhysRevE.92.042905} {\bibfield  {journal} {\bibinfo
  {journal} {Phys. Rev. E}\ }\textbf {\bibinfo {volume} {92}},\ \bibinfo
  {pages} {042905} (\bibinfo {year} {2015}{\natexlab{b}})}\BibitemShut
  {NoStop}%
\bibitem [{\citenamefont {Siemens}\ and\ \citenamefont
  {Schmelcher}(2020)}]{siemens2020}%
  \BibitemOpen
  \bibfield  {author} {\bibinfo {author} {\bibfnamefont {A.}~\bibnamefont
  {Siemens}}\ and\ \bibinfo {author} {\bibfnamefont {P.}~\bibnamefont
  {Schmelcher}},\ }\bibfield  {title} {\bibinfo {title} {Tunable order of
  helically confined charges},\ }\href
  {https://doi.org/10.1103/PhysRevE.102.012147} {\bibfield  {journal} {\bibinfo
   {journal} {Phys. Rev. E}\ }\textbf {\bibinfo {volume} {102}},\ \bibinfo
  {pages} {012147} (\bibinfo {year} {2020})}\BibitemShut {NoStop}%
\bibitem [{\citenamefont {Plettenberg}\ \emph {et~al.}(2017)\citenamefont
  {Plettenberg}, \citenamefont {Stockhofe}, \citenamefont {Zampetaki},\ and\
  \citenamefont {Schmelcher}}]{plettenberg2017}%
  \BibitemOpen
  \bibfield  {author} {\bibinfo {author} {\bibfnamefont {J.}~\bibnamefont
  {Plettenberg}}, \bibinfo {author} {\bibfnamefont {J.}~\bibnamefont
  {Stockhofe}}, \bibinfo {author} {\bibfnamefont {A.~V.}\ \bibnamefont
  {Zampetaki}},\ and\ \bibinfo {author} {\bibfnamefont {P.}~\bibnamefont
  {Schmelcher}},\ }\bibfield  {title} {\bibinfo {title} {Local equilibria and
  state transfer of charged classical particles on a helix in an electric
  field},\ }\href {https://doi.org/10.1103/PhysRevE.95.012213} {\bibfield
  {journal} {\bibinfo  {journal} {Phys. Rev. E}\ }\textbf {\bibinfo {volume}
  {95}},\ \bibinfo {pages} {012213} (\bibinfo {year} {2017})}\BibitemShut
  {NoStop}%
\bibitem [{\citenamefont {Siemens}\ and\ \citenamefont
  {Schmelcher}(2021)}]{siemens2021}%
  \BibitemOpen
  \bibfield  {author} {\bibinfo {author} {\bibfnamefont {A.}~\bibnamefont
  {Siemens}}\ and\ \bibinfo {author} {\bibfnamefont {P.}~\bibnamefont
  {Schmelcher}},\ }\bibfield  {title} {\bibinfo {title} {External-field-induced
  dynamics of a charged particle on a closed helix},\ }\href
  {https://doi.org/10.1103/PhysRevE.103.052217} {\bibfield  {journal} {\bibinfo
   {journal} {Phys. Rev. E}\ }\textbf {\bibinfo {volume} {103}},\ \bibinfo
  {pages} {052217} (\bibinfo {year} {2021})}\BibitemShut {NoStop}%
\bibitem [{\citenamefont {Gloy}\ \emph {et~al.}(2022)\citenamefont {Gloy},
  \citenamefont {Siemens},\ and\ \citenamefont {Schmelcher}}]{gloy2022}%
  \BibitemOpen
  \bibfield  {author} {\bibinfo {author} {\bibfnamefont {J.~F.}\ \bibnamefont
  {Gloy}}, \bibinfo {author} {\bibfnamefont {A.}~\bibnamefont {Siemens}},\ and\
  \bibinfo {author} {\bibfnamefont {P.}~\bibnamefont {Schmelcher}},\ }\bibfield
   {title} {\bibinfo {title} {Driven toroidal helix as a generalization of the
  {{Kapitza}} pendulum},\ }\href {https://doi.org/10.1103/PhysRevE.105.054204}
  {\bibfield  {journal} {\bibinfo  {journal} {Phys. Rev. E}\ }\textbf {\bibinfo
  {volume} {105}},\ \bibinfo {pages} {054204} (\bibinfo {year}
  {2022})}\BibitemShut {NoStop}%
\bibitem [{\citenamefont {Zampetaki}\ \emph {et~al.}(2018)\citenamefont
  {Zampetaki}, \citenamefont {Stockhofe},\ and\ \citenamefont
  {Schmelcher}}]{zampetaki2018}%
  \BibitemOpen
  \bibfield  {author} {\bibinfo {author} {\bibfnamefont {A.~V.}\ \bibnamefont
  {Zampetaki}}, \bibinfo {author} {\bibfnamefont {J.}~\bibnamefont
  {Stockhofe}},\ and\ \bibinfo {author} {\bibfnamefont {P.}~\bibnamefont
  {Schmelcher}},\ }\bibfield  {title} {\bibinfo {title} {Electrostatic bending
  response of a charged helix},\ }\href
  {https://doi.org/10.1103/PhysRevE.97.042503} {\bibfield  {journal} {\bibinfo
  {journal} {Phys. Rev. E}\ }\textbf {\bibinfo {volume} {97}},\ \bibinfo
  {pages} {042503} (\bibinfo {year} {2018})}\BibitemShut {NoStop}%
\bibitem [{\citenamefont {Zampetaki}\ \emph {et~al.}(2017)\citenamefont
  {Zampetaki}, \citenamefont {Stockhofe},\ and\ \citenamefont
  {Schmelcher}}]{zampetaki2017}%
  \BibitemOpen
  \bibfield  {author} {\bibinfo {author} {\bibfnamefont {A.~V.}\ \bibnamefont
  {Zampetaki}}, \bibinfo {author} {\bibfnamefont {J.}~\bibnamefont
  {Stockhofe}},\ and\ \bibinfo {author} {\bibfnamefont {P.}~\bibnamefont
  {Schmelcher}},\ }\bibfield  {title} {\bibinfo {title} {Pinned-to-sliding
  transition and structural crossovers for helically confined charges},\ }\href
  {https://doi.org/10.1103/PhysRevE.95.022205} {\bibfield  {journal} {\bibinfo
  {journal} {Phys. Rev. E}\ }\textbf {\bibinfo {volume} {95}},\ \bibinfo
  {pages} {022205} (\bibinfo {year} {2017})}\BibitemShut {NoStop}%
\bibitem [{\citenamefont {Zampetaki}\ \emph {et~al.}(2013)\citenamefont
  {Zampetaki}, \citenamefont {Stockhofe}, \citenamefont {Kr{\"o}nke},\ and\
  \citenamefont {Schmelcher}}]{zampetaki2013}%
  \BibitemOpen
  \bibfield  {author} {\bibinfo {author} {\bibfnamefont {A.~V.}\ \bibnamefont
  {Zampetaki}}, \bibinfo {author} {\bibfnamefont {J.}~\bibnamefont
  {Stockhofe}}, \bibinfo {author} {\bibfnamefont {S.}~\bibnamefont
  {Kr{\"o}nke}},\ and\ \bibinfo {author} {\bibfnamefont {P.}~\bibnamefont
  {Schmelcher}},\ }\bibfield  {title} {\bibinfo {title} {Classical scattering
  of charged particles confined on an inhomogeneous helix},\ }\href
  {https://doi.org/10.1103/PhysRevE.88.043202} {\bibfield  {journal} {\bibinfo
  {journal} {Phys. Rev. E}\ }\textbf {\bibinfo {volume} {88}},\ \bibinfo
  {pages} {043202} (\bibinfo {year} {2013})}\BibitemShut {NoStop}%
\bibitem [{\citenamefont {Reitz}\ and\ \citenamefont
  {Rauschenbeutel}(2012)}]{reitz2012}%
  \BibitemOpen
  \bibfield  {author} {\bibinfo {author} {\bibfnamefont {D.}~\bibnamefont
  {Reitz}}\ and\ \bibinfo {author} {\bibfnamefont {A.}~\bibnamefont
  {Rauschenbeutel}},\ }\bibfield  {title} {\bibinfo {title} {Nanofiber-{{Based
  Double-Helix Dipole Trap}} for {{Cold Neutral Atoms}}},\ }\href
  {https://doi.org/10.1016/j.optcom.2012.06.034} {\bibfield  {journal}
  {\bibinfo  {journal} {Opt. Comm.}\ }\textbf {\bibinfo {volume} {285}},\
  \bibinfo {pages} {4705} (\bibinfo {year} {2012})}\BibitemShut {NoStop}%
\bibitem [{\citenamefont {Willitsch}(2012)}]{willitsch2012}%
  \BibitemOpen
  \bibfield  {author} {\bibinfo {author} {\bibfnamefont {S.}~\bibnamefont
  {Willitsch}},\ }\bibfield  {title} {\bibinfo {title} {Coulomb-crystallised
  molecular ions in traps: Methods, applications, prospects},\ }\href
  {https://doi.org/10.1080/0144235X.2012.667221} {\bibfield  {journal}
  {\bibinfo  {journal} {Int. Rev. Phys. Chem.}\ }\textbf {\bibinfo {volume}
  {31}},\ \bibinfo {pages} {175} (\bibinfo {year} {2012})}\BibitemShut
  {NoStop}%
\end{thebibliography}%

\end{document}